# Theoretical Analysis of Hanbury Brown and Twiss Interferometry at Soft X-ray Free-Electron Lasers

Ivan A. Vartanyants[1,2,*] and Ruslan Khubbutdinov[1,2]

[1]*Deutsches Elektronen-Synchrotron DESY, Notkestraße 85, D-22607 Hamburg, Germany;*
[2]*National Research Nuclear University MEPhI (Moscow Engineering Physics Institute), Kashirskoe shosse 31, 115409 Moscow, Russia*

April 9, 2021

## Abstract

In this work we provide theoretical background for the analysis of second-order correlation functions in experiments performed at the soft x-ray free-electron lasers (XFELs). Typically, soft x-ray beamlines at XFELs are equipped by the variable line spacing (VLS) monochromators. We perform examination of the beam propagation through such VLS monochromator taking specially into account the interplay between the finite monochromator resolution and the exit slits width. We then provide general analysis of the second-order correlation intensities in spectral and spatial domains. Finally, we connect these functions with the statistical properties of the beam incoming to the monochromator unit in the limit of Gaussian Schell-model pulses.

[*]Corresponding author: Ivan.Vartaniants@desy.de

# I. Introduction

X-ray free-electron lasers (XFELs) in the soft and hard x-ray range are modern large-scale facilities that provide unique femtosecond ultrabright pulses of x-ray beams [1-7]. Besides the unique timing structure, XFELs are the sources with a high degree of spatial coherence [8-11]. The best way to analyze statistical properties of XFELs is to perform second-order intensity correlation analysis. In their pioneering experiment Hanbury Brown and Twiss (HBT) [12,13] measured intensities arriving from Sirius at two spatially separated telescopes and then correlated the corresponding signals. Such measurements can be performed both in spatial or time/frequency domains. They allowed to determine bunching/antibunching effects in optics [14] and led to the foundation of the quantum optics by Glauber [15].

A series of HBT experiments were performed nowadays at the soft and hard XFELs [16-22]. These experiments are especially spectacular at soft XFEL sources equipped by variable line spacing (VLS) monochromators [1,4,6] as soon as opening or closing the exit slits of the monochromator provide the knob to vary the spectral width of the x-ray radiation passing through these slits. Such measurements provide the way to estimate the average pulse duration in the specific conditions of XFEL operation [16-18,21,22] and provide important information on the beam statistics [19]. This may be especially important for experiments exploiting time resolution of the XFELs or utilizing their coherence properties and may yield an important feedback to accelerator scientists.

In this work, we present theoretical analysis of propagation of x-ray pulses through the VLS monochromator and then further to the pixelated detector, where spatial HBT measurements are, typically, performed. The position of this detector is usually considered in such a way that it will be sufficient resolution to resolve individual spatial modes of the incoming x-ray pulses. We specially consider the interplay between the monochromator resolution and the size of the exit slits of the monochromator. Importantly, we demonstrate relationship between the beam statistics determined in HBT interferometry with the statistical properties of the x-ray beam incoming to the VLS monochromator. In the end, we provide results for the Gaussian Schell-model pulses describing statistical properties of the XFELs.

We start in the next section with the basic analysis of x-ray propagation through the VLS monochromator.

# II. Propagation of the x-ray beams through the VLS grating

We consider the incoming field in the form

$$E'_{in}(\mathbf{r}, t) = E_{in}(\mathbf{r}, t)e^{i\mathbf{k_0}\mathbf{r}-i\omega_0 t}, \tag{1}$$

where $\mathbf{k_0} = k_0 \mathbf{s_0}$, $k_0 = \omega_0/c$, $\mathbf{s_0}$ is the direction of the incoming momentum vector, $\omega_0$ is the carrier frequency, and $c$ is the speed of light. In Eq. (1) we assume that the incoming amplitude $E_{in}(\mathbf{r}, t)$ is a slow varying function of its arguments. We can go to spatial-frequency space for the incoming amplitude $E_{in}(\mathbf{r}, t)$ by performing Fourier transform (FT)

$$E_{in}(\mathbf{r}, t) = \frac{1}{2\pi} \int_{-\infty}^{\infty} E_{in}(\mathbf{r}, \omega) e^{-i\omega t} d\omega. \tag{2}$$

If we substitute now Eq. (2) in Eq. (1) we see that the total frequency $\omega_t$ is defined as $\omega_t = \omega_0 + \omega$ and the region of convergence in frequency domain of the integral in Eq. (2) obeys inequality $\omega \ll \omega_0$.

We will now propagate each frequency $E_{in}(\mathbf{r}, \omega)$ separately through the beamline and, in the end, we will determine $E_f(\mathbf{r}, \omega)$ that is the final amplitude in spatial-frequency space. By performing FT we obtain in the spatial-time domain

$$E_f(\mathbf{r}, t) = \frac{1}{2\pi} \int_{-\infty}^{\infty} E_f(\mathbf{r}, \omega) e^{-i\omega t} d\omega. \tag{3}$$

Now calculating intensity

$$I_f(\mathbf{r}, t) = |E_f(\mathbf{r}, t)|^2 = \frac{1}{(2\pi)^2} \iint_{-\infty}^{\infty} E_f^*(\mathbf{r}, \omega) E_f(\mathbf{r}, \omega') e^{i\omega t - i\omega' t} d\omega d\omega' \tag{4}$$

and integrating this expression over the time gives $\int_{-\infty}^{\infty} exp[i(\omega - \omega')t] dt = (2\pi)\delta(\omega - \omega')$. Finally, we obtain for the integrated intensity at the position $\mathbf{r}$ [23]

$$I_f(\mathbf{r}) = \int_{-\infty}^{\infty} |E_f(\mathbf{r}, t)|^2 dt = \frac{1}{2\pi} \int_{-\infty}^{\infty} |E_f(\mathbf{r}, \omega)|^2 d\omega. \tag{5}$$

Expression (5) is stating the fact that the energy of the beam in time and frequency domains is conserved.

We will now consider scattering of the incoming field in the form of Eqs. (1-2) on a VLS grating in reflection geometry as depicted in Fig. 1(a) and will independently propagate each amplitude $E_{in}(\mathbf{r}, \omega)$ through this grating. It is well known that VLS gratings are characterized by changing the groove spacing according to

$$d(y) = d_0 + d_1 y + d_2 y^2 + \cdots, \tag{6}$$

where $d_0$ is the spacing at the pole of the grating located at the coordinate $y=0$ and $d_1$ and $d_2$ are the parameters for the variation of the line spacing with the coordinate $y$. It is also known (see, for example, [24]) that, if in the expansion (6) only first two terms are kept, such grating will focus the incoming beam at the focal distance

$$f = \left[\lambda d_1 \big/ d_0^2 \sin^2 \theta_f \right]^{-1}, \quad (7)$$

where $\theta_f$ is the exit scattering angle (see Fig. 1(a)). It is clear, that such VLS grating may be substituted by a plane grating with a spacing $d_0$ followed by the "virtual" lens with the focal distance given by the expression (7) (see Fig. 1(b)). In the following, we will consider this geometry for description of scattering on the VLS grating.

For the scattered field $E_{gr}(\mathbf{r}, \omega)$ from such a VLS grating one can write

$$E_{gr}(\mathbf{r}, \omega) = E_0 \int R_{gr}(\mathbf{r}')T_f(\mathbf{r}')P_f(\mathbf{r} - \mathbf{r}')E_{in}(\mathbf{r}', \omega)e^{i\mathbf{k}_0\mathbf{r}'}d\mathbf{r}', \quad (8)$$

where $E_0$ is the amplitude of the field in which we will further incorporate all not essential pre-integral factors. Here and below we assume that integration is performed from minus to plus infinity, where it is not specificly indicated. In Eq. (8) $R_{gr}(\mathbf{r})$ is the reflection function of the grating, $T_f(\mathbf{r})$ is the transmission function of the "virtual" lens given by the following expression

$$T_f(\mathbf{r}) = exp\left[-i\frac{k}{2f}r^2\right] \quad (9)$$

and

$$P_f(\mathbf{r} - \mathbf{r}') = \left(\frac{-ik_0}{2\pi f}\right) exp\left[i\frac{k}{2f}(\mathbf{r} - \mathbf{r}')^2\right] \quad (10)$$

is the propagator, written in the paraxial approximation for the focal distance $f$.

We will consider now, that the plane grating is located in the origin of the coordinate system as shown in Fig. 1. We consider the grooves along the $x'$-axis. In this geometry $R_{gr}(\mathbf{r}') = R_{gr}(y')$. We will also assume that the "virtual" lens in Eq. (9) is focusing only in one $y$-direction (as shown in Fig. 1(b)) $T_f(\mathbf{r}') = T_f(y')$ and that the observation plane is located at the focal plane of this lens. In this case in Eq. (8), we have for the product $T_f(y')P_f(\mathbf{r} - \mathbf{r}')exp(k_0 s_0^y \mathbf{r}')$ the following expression

$$T_f(y')P_f(\mathbf{r} - \mathbf{r}')e^{i\mathbf{k}_0\mathbf{r}'} = \left(\frac{-ik_0}{2\pi f}\right)^{1/2} exp\left[i\left(\frac{k}{2f}\right)y^2 - iq_y y'\right] P_f(x - x'), \quad (11)$$

where $q_y = ks^y - k_0 s_0^y$, $s^y$ is the $y$-component $s^y = y/f$ of the unit vector $\mathbf{s}$ towards the observation point, and $P_f(x - x')$ is the propagator in the x-direction. In Eq. (11) we considered that the x-component of the unit vector $s_0^x = 0$ (see Fig. 1(a)). Substituting all this in expression (8), we obtain

$$E_{gr}(x, q_y, \omega) = E_0 \iint R_{gr}(y') P_f(x - x') E_{in}(x', y', \omega) e^{-i q_y \cdot y'} dx' dy' . \qquad (12)$$

In this equation we can perform integration over x' that would lead to the following expression for the scattered amplitude

$$E_{gr}(x, q_y, \omega) = E_0 \int R_{gr}(y') E_{in}(x, y', \omega) e^{-i q_y \cdot y'} dy' , \qquad (13)$$

where

$$E_{in}(x, y', \omega) = \int P_f(x - x') E_{in}(x', y', \omega) dx' . \qquad (14)$$

We can now introduce the finite size grating function as

$$R_{gr}(y') = \sum_{n=-N/2}^{N/2} r_{gr}(y' - y_n) = R_N(y') \sum_{n=-\infty}^{\infty} r_{gr}(y' - y_n)$$

$$= R_N(y') \sum_{h_n=-\infty}^{\infty} r_{gr}(h_n) e^{i h_n y'} , \qquad (15)$$

where $N$ is the number of grating periods, $r_{gr}(y)$ is the reflection function of one period, and $y_n = d_0 \cdot n$, and $n = 0, \pm 1, \pm 2 \ldots$. In Eq. (15) $R_N(y)$ is a finite window of the size of the grating, that can be represented as

$$R_N(y') = rect\left(y'/D_N\right), \qquad (16)$$

where $D_N = d_0 \cdot N$ and $rect(y)$ is the rectangular function defined as

$$rect(y) = \begin{cases} 1, |y| \leq 1/2 \\ 0, |y| > 1/2 \end{cases}. \qquad (17)$$

In Eq. (15) we used also decomposition of the infinite periodic function over reciprocal space with $h_n = (2\pi/d_0) \cdot n$, and introduced the FT of the reflection function of one grating period as

$$r_{gr}(h_n) = \frac{1}{d}\int_{-d/2}^{d/2} r_{gr}(y)e^{-ih_n y}dy. \qquad (18)$$

Substituting now Eq. (15) in Eq. (13) and changing the order of summation and integration we obtain

$$E_{gr}(x, q_y, \omega) = E_0 \sum_{h_n=-\infty}^{\infty} r_{gr}(h_n) \int R_N(y')E_{in}(x,y',\omega)e^{-i(q_y-h_n)\cdot y'}dy'. \qquad (19)$$

Assuming now that the incoming amplitude is constant over the grating in the vertical direction $E_{in}(x,y',\omega) \cong E_{in}(x,\omega)$ we obtain from Eq. (19)

$$E_{gr}(x, \omega) = E_0 \sum_{h_n=-\infty}^{\infty} r_{gr}(h_n) E_{in}(x,\omega) \int R_N(y')e^{-i(q_y-h_n)\cdot y'}dy'. \qquad (20)$$

Performing now the integration

$$\int R_N(y')e^{-i(q_y-h_n)\cdot y'}dy' = \left[(q_y-h_n)\frac{D_N}{2}\right]^{-1}\sin\left[(q_y-h_n)\frac{D_N}{2}\right] = sinc[\alpha_n], \qquad (21)$$

where

$$\alpha_n = (q_y - h_n)D_N/2 \qquad (22)$$

we finally obtain for the amplitude of the field scattered from the grating

$$E_{gr}(x,\omega) = E_0 \sum_{h_n=-\infty}^{\infty} r_{gr}(h_n)\, sinc\,[\alpha_n]E_{in}(x,\omega). \qquad (23)$$

Now, considering scattering into one of the grating orders *n*, we get for the scattered amplitude

$$E_{gr}(x,\omega) = E_0 r_{gr}(h_n)\, sinc\,[\alpha_n]E_{in}(x,\omega). \qquad (24)$$

We will now determine the argument $\alpha_n$ of the *sinc*-function that can be presented as

$$\alpha_n = (q_y - h_n)D_N/2 = (ks^y - k_0 s_0^y - h_n)D_N/2 \\ = (k\cos\theta_f - k_0\cos\theta_i - h_n)D_N/2. \qquad (25)$$

The maximum of the *sinc*-function is at $\alpha_n=0$ that gives for the central frequency $\omega_0$ (we remind that $k_0=\omega_0/c$) the following condition (that is well known from optics the grating equation [25])

$$k_0(\cos\theta_f^n - \cos\theta_i) = h_n, \qquad (26)$$

where $\theta_i$ and $\theta_f^n$ are the incidence and scattered angles from the VLS grating as shown in Fig. 1(a). Now, taking into account that $k = (\omega_0 + \omega)/c$ and $\theta_f = \theta_f^n + \theta$, where $\omega \ll \omega_0$ and $\theta \ll \theta_f^n$, we substitute these relations in Eq. (25) and, neglecting small terms of the 2$^{nd}$ order, we obtain for the parameter $\alpha_n$

$$\alpha_n = \cos\theta_f^n \left[\left(\frac{\omega}{\omega_0}\right) - \theta\tan\theta_f^n\right]\frac{(k_0 D_N)}{2}. \qquad (27)$$

Condition $\alpha_n=0$ in Eq. (25) should be also valid for the other frequencies and corresponding angles and we obtain from this equation

$$k\cos\theta_f - k_0\cos\theta_i = h_n. \qquad (28)$$

Performing similar expansion as before $k = (\omega_0 + \omega')/c$ and $\theta_f = \theta_f^n + \theta$, where we introduced frequency ω' in the observation plane, we obtain

$$\left(\frac{\omega'}{\omega_0}\right) = \theta\tan\theta_f^n. \qquad (29)$$

Substituting this result in Eq. (27) we, finally, obtain for the parameter $\alpha_n$

$$\alpha_n = \cos\theta_f^n \left(\frac{\omega - \omega'}{\omega_0}\right)\frac{(k_0 D_N)}{2}. \qquad (30)$$

To determine an expression for the intensity in the observation plane, we substitute the scattered amplitude value in Eq. (24) to Eq. (5) and obtain

$$I_{gr}(x, \omega') = |E_0|^2 |r_{gr}(h_n)|^2 \int R(\omega' - \omega)|E_{in}(x,\omega)|^2 d\omega, \qquad (31)$$

where we introduced the function $R(\omega'-\omega)$ as

$$R(\omega' - \omega) = sinc^2(\alpha_n). \qquad (32)$$

We will show now that this function is, in fact, a resolution function of the monochromator unit. Indeed, if we consider, that the incoming field is monochromatic, with a fixed frequency $\omega_0$, we can write the incoming field as $E_{in}(x, \omega) \to (2\pi)\delta(\omega - \omega_0)E_{in}(x)$. Substituting this expression in Eq. (31) gives

$$I_{gr}(x, \omega') = |E_0|^2 |r_{gr}(h_n)|^2 R(\omega' - \omega_0)|E_{in}(x)|^2 \qquad (33)$$

that shows that the function $R(\omega' - \omega_0)$ gives the broadening of the monochromatic frequency $\omega_0$ and may be treated as resolution function. The full-width half-maximum (FWHM) of the resolution function $R(\omega'-\omega)$ defines, typically, the resolution of the VLS grating. The resolution function of the monochromator is often considered to be a Gaussian function

$$R(\omega) = exp\left(-\omega^2/2\sigma_r^2\right), \qquad (34)$$

where $\sigma_r$ is the root mean square (rms) value of the resolution function.

In a typical experiment at an XFEL in order to obtain spectral characteristics of the pulse the measured two-dimensional intensity distribution $I_{gr}(x,\omega)$ is integrated over the *x*-axis thus providing the one-dimensional spectrum $I_{gr}(\omega)$. Performing such integration in Eq. (31) and also taking into account Eq. (14) we obtain for the spectrum

$$I_{gr}(\omega) = \int I_{gr}(x,\omega)dx =$$
$$= |E_0|^2 |r_{gr}(h_n)|^2 \iint R(\omega-\omega')|E_{in}(x',\omega')|^2 d\omega' dx', \qquad (35)$$

In this derivation, we considered the following property of the propagator [26]

$$\int P_f^*(x-x')P_f(x-x'')dx = \delta(x'-x''). \qquad (36)$$

Equation (35) provides an expression for the measurements of the single pulse spectrum at the XFEL experiment by the VLS grating spectrometer expressed through the incoming beam intensity $|E_{in}(x',\omega')|^2$.

In the next section we determine which kind of information may be obtained by performing HBT analysis in the spectral domain and relate it to the statistical properties of the beam incoming on the monochromator unit.

### III. Second-order correlations in the frequency domain

We can now evaluate the second-order correlation functions in frequency domain according to its definition as

$$g^{(2)}(\omega_1,\omega_2) = \frac{\langle I_{gr}(\omega_1)I_{gr}(\omega_2)\rangle}{\langle I_{gr}(\omega_1)\rangle\langle I_{gr}(\omega_2)\rangle}, \qquad (37)$$

where intensities $I_{gr}(\omega)$ are defined in Eq. (35). Substituting these intensities in Eq. (37), we obtain

$$g^{(2)}(\omega_1,\omega_2) =$$
$$= \frac{\int R(\omega_1-\omega')R(\omega_2-\omega'')\langle|E_{in}(x',\omega')|^2|E_{in}(x'',\omega'')|^2\rangle dx'dx''d\omega'd\omega''}{\int R(\omega_1-\omega')\langle|E_{in}(x',\omega')|^2\rangle d\omega'dx' \int R(\omega_2-\omega')\langle|E_{in}(x',\omega')|^2\rangle d\omega'dx'}. \qquad (38)$$

Assuming Gaussian statistics for the incoming field, we have (see, for example, [27])

$$\langle |E_{in}(x',\omega')|^2 |E_{in}(x'',\omega'')|^2 \rangle$$
$$= \langle |E_{in}(x',\omega')|^2 \rangle \langle |E_{in}(x'',\omega'')|^2 \rangle + |\langle E_{in}^*(x',\omega') E_{in}(x'',\omega'') \rangle|^2 \quad (39)$$
$$= S_{in}(x',\omega') S_{in}(x'',\omega'') + |W_{in}(x',x'',\omega',\omega'')|^2 ,$$

where we have introduced the spectral density as

$$S_{in}(x,\omega) = \langle |E_{in}(x,\omega)|^2 \rangle \quad (40)$$

and cross-spectral density as

$$W_{in}(x',x'',\omega',\omega'') = \langle E_{in}^*(x',\omega') E_{in}(x'',\omega'') \rangle \quad (41)$$

Substituting Eq. (39) in Eq. (38) we obtain

$$g^{(2)}(\omega_1,\omega_2)$$
$$= 1 + \frac{\int R(\omega_1 - \omega') R(\omega_2 - \omega'') |W_{in}(x',x'',\omega',\omega'')|^2 dx' dx'' d\omega' d\omega''}{\int R(\omega_1 - \omega') S_{in}(x',\omega') dx' d\omega' \int R(\omega_2 - \omega') S_{in}(x',\omega') dx' d\omega'} . \quad (42)$$

Assuming now cross-spectral purity of the incoming x-ray beam, we have

$$W_{in}(x',x'',\omega',\omega'') = W_{in}(\omega',\omega'') W_{in}(x',x'') \quad (43)$$

and

$$S_{in}(x',\omega') = S_{in}(\omega') S_{in}(x') . \quad (44)$$

Substituting these expressions in Eq. (42), we, finally, obtain

$$g^{(2)}(\omega_1,\omega_2) = 1 + \varsigma_{in} g_{in}(\omega_1,\omega_2) , \quad (45)$$

where

$$\varsigma_{in} = \frac{\iint |W_{in}(x',x'')|^2 dx' dx''}{[\int S_{in}(x) dx]^2} \quad (46)$$

and

$$g_{in}(\omega_1,\omega_2) = \frac{\iint R(\omega_1 - \omega') R(\omega_2 - \omega'') |W_{in}(\omega',\omega'')|^2 d\omega' d\omega''}{\int R(\omega_1 - \omega') S_{in}(\omega') d\omega' \int R(\omega_2 - \omega') S_{in}(\omega') d\omega'} . \quad (47)$$

By that, we expressed the second-order correlation function $g^{(2)}(\omega_1,\omega_2)$ measured in the frequency domain through the statistical properties of the x-ray radiation incoming on the monochromator unit. We immediately see, that the contrast $\varsigma_{in}$ defined in Eq. (46) is determined by the degree of spatial coherence of the incoming x-ray beam [28].

We have to recall here that the spectral density in spatial $W_{in}(x_1,x_2)$ and frequency $W_{in}(\omega_1,\omega_2)$ domains is defined through its first-order correlation functions as

$$W_{in}(x_1, x_2) = [S_{in}(x_1)]^{1/2}[S_{in}(x_2)]^{1/2} g_{in}^{(1)}(x_1, x_2),$$
$$W_{in}(\omega_1, \omega_2) = [S_{in}(\omega_1)]^{1/2}[S_{in}(\omega_2)]^{1/2} g_{in}^{(1)}(\omega_1, \omega_2),$$
(48)

where the first-order correlation functions $g_{in}^{(1)}(x_1, x_2)$ and $g_{in}^{(1)}(\omega_1, \omega_2)$ are defined as

$$g_{in}^{(1)}(x_1, x_2) = \frac{\langle E_{in}^*(x_1) E_{in}(x_2) \rangle}{\sqrt{\langle |E_{in}(x_1)|^2 \rangle}\sqrt{\langle |E_{in}(x_2)|^2 \rangle}},$$
$$g_{in}^{(1)}(\omega_1, \omega_2) = \frac{\langle E_{in}^*(\omega_1) E_{in}(\omega_2) \rangle}{\sqrt{\langle |E_{in}(\omega_1)|^2 \rangle}\sqrt{\langle |E_{in}(\omega_2)|^2 \rangle}}.$$
(49)

We can now further analyze an expression for the correlation function $g_{in}(\omega_1, \omega_2)$ in Eq. (47). We note, that resolution of the monochromator is typically much narrower then the bandwidth of the incoming radiation. Assuming perfect monochromator resolution or substituting resolution function by a delta-function $R(\omega_1 - \omega') \to \delta(\omega_1 - \omega')$, we obtain for the correlation function $g_{in}(\omega_1, \omega_2)$ in Eq. (47)

$$g_{in}(\omega_1, \omega_2) = \frac{|W_{in}(\omega_1, \omega_2)|^2}{S_{in}(\omega_1) S_{in}(\omega_2)} = \left| g_{in}^{(1)}(\omega_1, \omega_2) \right|^2 \quad (50)$$

and we obtain for the second-order correlation function in Eq. (45) the following expression (that is similar to known Siegert relation in time domain [29])

$$g^{(2)}(\omega_1, \omega_2) = 1 + \varsigma_{in} \left| g_{in}^{(1)}(\omega_1, \omega_2) \right|^2. \quad (51)$$

By this expression the $g^{(2)}$-function in spectral domain is expressed through the square modulus of the first-order correlation function of the incoming beam. As we have seen from our analysis, this expression is valid only if one can assume an ideal resolution of the monochromator. It is interesting to note also, that for a chaotic source the second-order correlation function in spectral domain on maximum is $g_{max}^{(2)} = 1 + \varsigma_{in}$ (as soon as $g_{max}^{(1)}(\omega_1, \omega_2) = 1$ at $\omega_1=\omega_2=\omega$). We have seen that the contrast $\varsigma_{in}$ is directly related to the degree of spatial coherence of the incoming beam (see Eq. (45)) as such $g_{max}^{(2)} \leq 2$ and will be equal to two only for the fully coherent beam in spectral domain. We will see in the following how these results will be modified when the final resolution of the VLS monochromator will be taken into account.

In the next section we will analyze which kind of information may be obtained by performing the HBT interferometry in the spatial domain and will relate it to the statistical properties of the beam incoming on the VLS monochromator.

### IV. Second-order correlations in the spatial domain

The HBT experiments at the soft x-ray beamlines at the XFEL facilities in the spatial domain are performed on a pixelated detector positioned far from the focal plane of the beamline. Such arrangement provides sufficient sampling of the incoming beam at the detector that allows to resolve spatial modes of the XFEL individual pulses.

We introduce now a 2D detector with the coordinates $x_D, y_D$ at which single pulse intensities $I_D(x_D, y_D)$ from the XFEL are measured for the further correlation analysis (see Fig. 2). As soon as we measure intensities, they are connected with the amplitudes of the incoming field in spatial-frequency domain as (see Eq. (5))

$$I_D(x^D, y^D) = \frac{1}{2\pi} \int_{-\infty}^{\infty} |E_D(x^D, y^D, \omega)|^2 d\omega . \tag{52}$$

Each of these intensities is typically integrated in the vertical (dispersion) direction to provide a 1D distribution of the intensity

$$I_D(x^D) = \frac{1}{2\pi} \iint_{-\infty}^{\infty} |E_D(x^D, y^D, \omega)|^2 d\omega dy^D . \tag{53}$$

The basic idea of HBT interferometry is the correlation of intensities at different spatial positions, or, in other words, measurements of the second-order correlation functions. In such measurements the normalized second-order correlation function can be defined as

$$g^{(2)}(x_1^D, x_2^D) = \frac{< I_D(x_1^D) I_D(x_2^D) >}{< I_D(x_1^D) >< I_D(x_2^D) >}, \tag{54}$$

where averaging, denoted by brackets <...>, is performed over a large ensemble of different realizations of the wave field. In the HBT experiment for the nonstationary source, such as XFELs, averaging may be performed over different pulses, with the assumption that all pulses are realizations of the same statistical process.

Now, we will express intensities $I_D(x^D)$, as given in Eq. (53), through the amplitudes of the wavefield incoming on the monochromator unit and then will correlate these intensities according to Eq. (54). The x-ray field amplitudes are the result of propagation over the free space from the exit slits of the monochromator equipped by the VLS grating. Such slits are located in the focal plane of the VLS grating and the field amplitudes $E_D(x^D, y^D, \omega)$ may be presented as

$$E_D(x^D, y^D, \omega) = \iint P_L(x^D - x')P_L(y^D - y')E_{sl}(x', y', \omega)dx'dy', \qquad (55)$$

where $P_L(x^D - x')$ is the propagator (see Eq. (10)), the distance $L$ is defined in Fig. 2 as distance from the exit slits of the monochromator to the detector plane, and $E_{sl}(x,y)$ is the amplitude of the field passing through the exit slits, $x'$, $y'$ are the coordinates in the slits plane.

The amplitude of the field after the exit slits $E_{sl}(r,\omega)$ may be obtained by multiplying the amplitude of the field scattered from the VLS grating to a fixed order $E_{gr}(x,\omega)$ (see Eq. (24)) by the transmission function of the slits $T_{sl}(\omega)$

$$\begin{aligned} E_{sl}(x', y', \omega) &= T_{sl}(\omega)E_{gr}(x, \omega) \\ &= E_0 r_{gr}(h_n)T_{sl}(\omega)\mathrm{sinc}[\alpha_n] \int P_f(x' - x_1)E_{in}(x_1, \omega)dx_1, \end{aligned} \qquad (56)$$

where we took into account an expression (14) for the incoming amplitude $E_{in}(x, \omega)$. In Eq. (56), the transmission function of the slits $T_{sl}(\omega)$ is defined as a finite window in the spectral domain of the size of the spectral bandpass $D_\omega^{sl}$ as

$$T_{sl}(\omega) = rect\left[\frac{\omega}{D_\omega^{sl}}\right], \qquad (57)$$

where $rect(x)$ is the rectangular function defined in Eq. (17). Substituting now Eqs. (55, 56) in Eq. (53) we obtain for the intensity on the detector

$$I_D(x^D) = |E_0|^2|r_{gr}(h_n)|^2 \iiint_{-\infty}^{\infty} P_{f+L}^*(x^D - x_1)P_{f+L}(x^D - x_2)\tilde{T}_{sl}(\omega)E_{in}^*(x_1, \omega)E_{in}(x_2, \omega)dx_1 dx_2 d\omega, \qquad (58)$$

where we introduced a function $\tilde{T}_{sl}(\omega)$ that is defined as

$$\tilde{T}_{sl}(\omega) = \int T_{sl}^2(\omega')R(\omega - \omega')d\omega'. \qquad (59)$$

In deriving Eq. (58) we used the following properties of the propagator (see, for example, [26])

$$\int P_{L_2}(x' - x)P_{L_1}(x - x'')dx = P_{L_1+L_2}(x' - x'') \qquad (60)$$

as well as Eq. (36). According to the equation (59), the square of the exit slit transmission function $T_{sl}^2(\omega')$ defined in Eq. (57) is convoluted with the resolution function $R(\omega)$ given in Eq. (34). For the function $T_{sl}^2(\omega)$ in Eq. (59) the following relationship is valid $T_{sl}^2(\omega) = T_{sl}(\omega)$, due to its definition in Eq. (57).

We substitute now Eq. (58) for the intensity $I_D(x^D)$ to the expression (54) for the $g^{(2)}$-function and assume that the incoming x-ray radiation obeys Gaussian statistics. In this case (see, for example, [27])

$$\langle E_{in}^*(x_1,\omega_1)E_{in}(x_2,\omega_1)E_{in}^*(x_3,\omega_2)E_{in}(x_4,\omega_2)\rangle =$$
$$= \langle E_{in}^*(x_1,\omega_1)E_{in}(x_2,\omega_1)\rangle\langle E_{in}^*(x_3,\omega_2)E_{in}(x_4,\omega_2)\rangle \quad (61)$$
$$+ \langle E_{in}^*(x_1,\omega_1)E_{in}(x_4,\omega_2)\rangle\langle E_{in}^*(x_3,\omega_2)E_{in}(x_2,\omega_1)\rangle.$$

We further introduce the spectral density function and cross-spectral density function in spatial-frequency domain as given in Eqs. (40, 41) and, further, assume cross-spectral purity of the incoming x-ray radiation as in Eqs. (43, 44). After long, but straight forward calculations, one can, finally, obtain for the $g^{(2)}$-function

$$g^{(2)}(x_1^D, x_2^D) = 1 + \varsigma_{in}(D_\omega)|g_{in}(x_1^D, x_2^D)|^2. \quad (62)$$

In this expression, the contrast function $\varsigma_{in}(D_\omega)$, which depends on the radiation bandwidth $D_\omega$ is defined as

$$\zeta_{in}(D_\omega) = \frac{\iint \tilde{T}_{sl}(\omega_1)\tilde{T}_{sl}(\omega_2)|W_{in}(\omega_1,\omega_2)|^2 d\omega_1 d\omega_2}{[\int \tilde{T}_{sl}(\omega)S_{in}(\omega)d\omega]^2} \quad (63)$$

and a correlation function $g_{in}(x_1^D, x_2^D)$ is determined by

$$g_{in}(x_1^D, x_2^D) = \frac{\iint P_{f+L}^*(x_1^D - x_1)P_{f+L}(x_2^D - x_2)W_{in}(x_1,x_2)dx_1 dx_2}{[S_D(x_1^D)]^{1/2}[S_D(x_2^D)]^{1/2}}, \quad (64)$$

where

$$S_D(x_i^D) = \iint P_{f+L}^*(x_i^D - x_1)P_{f+L}(x_i^D - x_2)W_{in}(x_1,x_2)dx_1 dx_2 \quad (65)$$

and *i=1,2*.

By that, we expressed the second-order correlation function measured at the far detector (Eq. (62) through the statistical properties of x-ray radiation incoming on the VLS monochromator. We will show in the following that by changing the exit slits opening $D_\omega$ the contrast function $\varsigma_{in}(D_\omega)$ also changes according to Eq. (63). This, finally, allows to determine pulse duration of the incoming x-ray pulses.

In the next section, we will represent the incoming x-ray pulses by the Gaussian Schell-model.

## V. Gaussian Schell-model x-ray pulses

After receiving these general results both in the frequency and spatial domains we will consider now special type of the incoming pulsed beams in the form of the Gaussian Schell-model (GSM) pulses. In this case the cross-spectral density in the spatial-frequency domain can be written as [28,31,32]

$$W_{in}(x_1, x_2; \omega_1, \omega_2) = W_0 W(x_1, x_2) W(\omega_1, \omega_2), \quad (66)$$

where the spatial dependence is defined by the function

$$W(x_1, x_2) = exp\left[-\frac{x_1^2 + x_2^2}{4\sigma_I^2} - \frac{(x_2 - x_1)^2}{2l_c^2}\right], \quad (67)$$

where $\sigma_I$ is the rms size and $l_c$ is the spatial coherence of the incoming beam. The frequency dependence is defined by

$$W(\omega_1, \omega_2) = exp\left[-\frac{\omega_1^2 + \omega_2^2}{4\Omega^2} - \frac{(\omega_2 - \omega_1)^2}{2\Omega_c^2}\right], \quad (68)$$

where $\Omega$ is the rms spectral width of radiation, $\Omega_c$ is the spectral coherence, and we assumed that the frequency is counted from the corresponding grating order. The parameters $\Omega$ and $\Omega_c$ can be also expressed through the rms values of the pulse duration $\sigma_T$ and coherence time $\tau_c$ of the pulse in front of the monochromator unit [31,32]

$$\Omega^2 = \frac{1}{\tau_c^2} + \frac{1}{4\sigma_T^2}, \quad \Omega_c = \frac{\tau_c}{\sigma_T}\Omega. \quad (69)$$

For the SASE pulses at XFELs for most of cases the coherence time $\tau_c$ in front of the monochromator unit is much shorter than the pulse duration $\sigma_T$ ($\tau_c \ll \sigma_T$). Taking this into account we have from Eqs. (69)

$$\Omega \approx \frac{1}{\tau_c}, \quad \Omega_c \approx \frac{1}{\sigma_T}. \quad (70)$$

In this limit, we would have for the spectral density

$$S_{in}(\omega) = S_0 exp\left[-\frac{\tau_c^2 \omega^2}{2}\right] \quad (71)$$

and for the first-order correlation function

$$g_{in}^{(1)}(\omega_2 - \omega_1) = exp\left[-\frac{\sigma_T^2(\omega_2 - \omega_1)^2}{2}\right]. \tag{72}$$

Here, we would like to note, that if one can neglect by the monochromator resolution than Eq. (51) will be valid. With combination of Eq. (72) this allows to determine the average pulse duration through analysis of the $g^{(2)}$-function as a function of $\Delta\omega$ in spectral domain.

Next, we will see which results may be obtained if we take into account the finite energy resolution of the monochromator unit. We will consider spectral and spatial domains separately in the following.

## VI.    Spectral domain in the frame of GSM

Now we can perform integration in Eqs. (46,47) assuming that the incoming beam obeys Gaussian Schell-model. Substituting in these equations expressions (66-68) for the incoming cross-spectral density function we obtain for the contrast

$$\varsigma_{in} = \left[1 + 4\left(\frac{\sigma_I}{l_c}\right)^2\right]^{-1/2} \tag{73}$$

and for the correlation function [33]

$$g_{in}(\omega_2 - \omega_1) = \frac{\alpha exp\left[-\frac{1}{\alpha\beta}\frac{(\omega_2 - \omega_1)^2}{\Omega_c^2}\right]}{(\alpha\beta)^{1/2}}, \tag{74}$$

where

$$\alpha = 1 + \left(\frac{\sigma_r}{\Omega}\right)^2, \beta = 1 + \left(\frac{\sigma_r}{\Omega}\right)^2\left[1 + 4\left(\frac{\Omega}{\Omega_c}\right)^2\right]. \tag{75}$$

Now, in the limit of Eqs. (70) we have

$$g_{in}(\omega_2 - \omega_1) = \frac{\alpha exp\left[-\frac{\sigma_T^2}{\alpha\beta}(\omega_2 - \omega_1)^2\right]}{(\alpha\beta)^{1/2}}, \tag{76}$$

where

$$\alpha = 1 + (\sigma_r \tau_c)^2 \,, \beta = 1 + (\sigma_r \tau_c)^2 \left[1 + 4\left(\frac{\sigma_T}{\tau_c}\right)^2\right]. \tag{77}$$

Below, we will analyse this expression. First, we immediately see, that if the monochromator has a perfect resolution $\sigma_r=0$, then both α and β are equal to one and $g_{in}(\omega_2 - \omega_1) = \left|g_{in}^{(1)}(\omega_2 - \omega_1)\right|^2$, where $g_{in}^{(1)}(\omega_2 - \omega_1)$ is defined by Eq. (72). We note, that this result coincides with our previous result of Eqs. (50,51).

Next, we want to note, that for typical SASE beams the product $\sigma_r\tau_c\ll 1$. In this case we have for the coefficients in Eq. (77) α≈1 and β≈$1 + 4(\sigma_r \sigma_T)^2$. Substituting this in Eq. (76) we obtain for $g_{in}(\omega_2-\omega_1)$

$$g_{in}(\omega_2 - \omega_1) = \frac{exp\left[-\frac{\sigma_T^2}{1 + 4\sigma_r^2\sigma_T^2}(\omega_2 - \omega_1)^2\right]}{(1 + 4\sigma_r^2\sigma_T^2)^{1/2}}. \tag{78}$$

We see from this expression that when we take into account the finite resolution of the monochromator $\sigma_r$, the function $g_{in}(\omega_2-\omega_1)$ does not reach unity even at $\omega_1=\omega_2$. Its maximum value at the same frequency values is equal to

$$g_{in}(0) = \frac{1}{(1 + 4\sigma_r^2\sigma_T^2)^{1/2}}. \tag{79}$$

Substituting this value of $g_{in}(\omega_2-\omega_1)$ in expression for $g^{(2)}$-function (45), and assuming that the resolution of the VLS monochromator is known we may determine the pulse duration $\sigma_T$ as

$$\sigma_T = \frac{1}{2\sigma_r}\left[\frac{\varsigma_{in}^2}{\left(g_{in}^{(2)}(\omega,\omega) - 1\right)^2} - 1\right]^{1/2}. \tag{80}$$

Finally, HBT analysis in spectral domain with the assumption that XFEL pulses obeys GSM provides the way to determine the pulse duration.

## VII. Spatial domain in the frame of GSM
### a) Contrast function

Now we will focus on evaluation of the contrast function $\zeta_{in}(D_\omega)$ in Eq. (63) and will show how an average pulse duration of XFEL pulses may be determined by evaluating this contrast function. Expressing the cross-spectral density in the frequency domain through its first-order correlation function (see Eq. (48)) we obtain for the contrast function in Eq. (63)

$$\zeta_{in}(D_\omega) = \frac{\iint_{-\infty}^{\infty} \tilde{T}_{sl}(\omega_1)\tilde{T}_{sl}(\omega_2)S_{in}(\omega_1)S_{in}(\omega_2)\left|g_{in}^{(1)}(\omega_1,\omega_2)\right|^2 d\omega_1 d\omega_2}{\left[\int_{-\infty}^{\infty} \tilde{T}_{sl}(\omega)S_{in}(\omega)d\omega\right]^2}. \qquad (81)$$

First, we assume that the spectral first-order correlation function is uniform or of the Schell type. This means that it depends only on the difference of frequencies as $g_{in}^{(1)}(\omega_1,\omega_2) = g_{in}^{(1)}(\omega_2 - \omega_1)$. In this case, after changing of the variables, we have for the nominator of the contrast function $\zeta_2(D_\omega)$ in Eq. (81) (see, for example, [30])

$$\iint_{-\infty}^{\infty} \tilde{T}_{sl}(\omega_1)\tilde{T}_{sl}(\omega_2)S_{in}(\omega_1)S(\omega_2)\left|g_{in}^{(1)}(\omega_2 - \omega_1)\right|^2 d\omega_1 d\omega_2$$
$$= \int_{-\infty}^{\infty} F(\omega)\left|g_{in}^{(1)}(\omega)\right|^2 d\omega, \qquad (82)$$

where $F(\omega)$ is the autocorrelation function

$$F(\omega) = \int_{-\infty}^{\infty} \tilde{\tilde{T}}_{sl}(\omega')\tilde{\tilde{T}}_{sl}(\omega' + \omega)\, d\omega', \qquad (83)$$

and

$$\tilde{\tilde{T}}_{sl}(\omega) = S_{in}(\omega)\tilde{T}_{sl}(\omega). \qquad (84)$$

Substituting these results in Eq. (81) we obtain for the contrast function $\zeta_{in}(D_\omega)$

$$\zeta_{in}(D_\omega) = \frac{\int_{-\infty}^{\infty} F(\omega)\left|g_{in}^{(1)}(\omega)\right|^2 d\omega}{\left[\int_{-\infty}^{\infty} \tilde{\tilde{T}}(\omega)d\omega\right]^2}. \qquad (85)$$

This is quite a general expression assuming only uniformity of the spectral first-order correlation function.

Next approximation can be made by the assumption that the bandwidth of x-ray radiation incoming to the monochromator unit is much wider than the transmitted one by the exit slits $D_\omega^{sl} \ll \Omega$. In this case we can also assume that $S_{in}(\omega)$ is constant in the integration region in Eq. (83), and can substitute $\tilde{\tilde{T}}_{sl}(\omega)$ by $\tilde{T}_{sl}(\omega)$ in this equation.

By substituting Eq. (72) in Eq. (85) and varying the bandwidth of x-ray radiation by opening and closing the exit slits a typical dependence of the contrast function can be obtained. Fitting this curve to the experimentally determined values of the contrast will give us the rms value of the pulse duration $\sigma_T$ and hence the average pulse duration $T=2.355 \cdot \sigma_T$.

We will now consider two limits for evaluation of the contrast function $\zeta_{in}(D_\omega)$. In the first limit, the opening of the slits will be much larger than the resolution function width $D_\omega^{sl} \gg \sigma_{res}$, in the second limit, the opposite will be assumed, $D_\omega^{sl} \ll \sigma_{res}$. In the first limit, the function $\tilde{T}_{sl}(\omega)$ in Eq. (59) may be, with a good approximation, substituted by the rectangular function $\tilde{T}_{sl}(\omega) \cong T_{sl}^2(\omega') = T_{sl}(\omega) = rect[\omega/D_\omega^{sl}]$. In this case, we have for the autocorrelation function $F(\omega)$ in Eq. (83) (see, for example, [30])

$$F(\omega) = \int_{-\infty}^{\infty} T_{sl}(\omega') T_{sl}(\omega' + \omega)\, d\omega = \begin{cases} 1 - \left|\omega/D_\omega^{sl}\right|, \omega \leq D_\omega^{sl} \\ 0, \omega > D_\omega^{sl} \end{cases}, \qquad (86)$$

Substituting this result, as well as the first-order correlation function from Eq. (72), in expression (85) for the contrast function $\zeta_{in}(D_\omega)$, we obtain

$$\zeta_{in}(D_\omega) = \frac{\int_{-D_\omega^{sl}}^{D_\omega^{sl}} [1 - |\omega/D_\omega^{sl}|]\, e^{-\sigma_T^2 \omega^2}\, d\omega}{[D_\omega^{sl}]^2}. \qquad (87)$$

Performing the integration, we obtain in this limit (see, for example, [16])

$$\zeta_{in}(D_\omega) = \frac{\sqrt{\pi}}{D_\omega^{sl}\sigma_T} erf(D_\omega^{sl}\sigma_T) + \frac{1}{\left(D_\omega^{sl}\sigma_T\right)^2}\left[e^{-\left(D_\omega^{sl}\sigma_T\right)^2} - 1\right], \qquad (88)$$

where $erf(x)$ is an error function.

According to definition of the coherence time for rectangular slits $\tau_c = 2\pi/D_\omega^{sl}$ and we obtain in the limit $\tau_c/\sigma_T \ll 1$ for the contrast function $\zeta_{in}(D_\omega) \sim [1/(2\sqrt{\pi})](\tau_c/\sigma_T) = [1/(2\sqrt{\pi})](1/M_t)$, where $M_t$ is the number of temporal modes.

In the other limit, $D_\omega^{sl} \ll \sigma_r$, according to Eq. (59), the function $\tilde{T}_{sl}(\omega)$ can be represented by the resolution function $R(\omega)$. Substituting an expression (34) for the resolution function $R(\omega)$ into Eqs. (83) and (85) one can show that the contrast function $\zeta_{in}(D_\omega)$ in this limit has the following form

$$\zeta_{in}(D_\omega) = \frac{1}{\sqrt{1 + 2(\sigma_r\sigma_T)^2}}. \qquad (89)$$

In this case of a Gaussian spectrum, coherence time is well approximated by $\tau_c = \sqrt{\pi}/\sigma_r$ and substituting it in expression (89) gives

$$\zeta_{in}(D_\omega) = \frac{1}{\sqrt{1 + 2\pi(\sigma_T/\tau_c)^2}} \ . \tag{90}$$

In the limit of $\tau_c/\sigma_T \gg 1$ we obtain for the contrast function $\zeta_{in}(D_\omega) \cong 1 - \pi(\sigma_T/\tau_c)^2$. Expressions (89) and (90) indicate that in the case of a limited resolution of the monochromator the contrast function is always below one. If the resolution of the monochromator is known, in the conditions $D_\omega^{sl} \ll \sigma_r$, we can get an estimate of the pulse duration from Eq. (89) as

$$\sigma_T = \frac{1}{\sigma_r}\sqrt{\frac{1 - \zeta_{in}^2(D_\omega)}{2\zeta_{in}^2(D_\omega)}} \ . \tag{91}$$

This equation provides an alternative (to spectral domain) way to determine the pulse duration of XFEL pulses in the conditions of GSM.

**b) Spatial correlations**

We now express the spatial part of the 2$^{nd}$-order correlation function given in Eq. (64) with an assumption that the incoming field is of the Gaussian Schell-model type. We will assume also that measurements are performed in the far-field and the propagators in Eqs. (64, 65) may be expressed by simple exponential functions

$$P_{f+L}(x^D - x) \propto e^{-iq^D x}, q^D = k\frac{x^D}{(f + L)} \ . \tag{92}$$

Substitution of these relations for the propagator in Eqs. (64, 65) gives for the correlation function

$$g_{in}(q_1^D, q_2^D) = \frac{\iint exp[iq_1^D x_1 - iq_2^D x_2]W_{in}(x_1, x_2)dx_1 dx_2}{[S_D(q_1^D)]^{1/2}[S_D(q_2^D)]^{1/2}}, \tag{93}$$

and for $S_D(q_i^D)$

$$S_D(q_i^D) = \int exp[iq_i^D x_1 - iq_i^D x_2]W_{in}(x_1, x_2)dx_1 dx_2 \ . \tag{94}$$

Performing direct integration in Eqs. (93, 94) with the cross-spectral density function given in Eqs. (66, 67) (see Appendix I for details) we obtain

$$g_{in}(q_1^D, q_2^D) = e^{-\beta(q_2^D - q_1^D)^2}, \tag{95}$$

where parameter $\beta$ is equal to

$$\beta = \frac{2\sigma_I^4}{l_c^2 + 4\sigma_I^2} \ . \tag{96}$$

Finally, we have for the second-order correlation function in Eq. (62) in the far field and for the GSM pulses for the incoming x-ray radiation the following expression

$$g^{(2)}(x_1^D, x_2^D) = 1 + \varsigma_{in}(D_\omega)e^{-2\beta(q_2^D-q_1^D)^2}, \qquad (97)$$

where the contrast function $\varsigma_{in}(D_\omega)$ is defined by the expression (85).

## VIII. Summary and outlook

In summary, we provided here theoretical description of x-ray propagation through the VLS monochromator in a typical HBT interferometry experiment. We demonstrated how second-order intensity measurements performed in such experiment are related to statistical properties of the x-ray pulses incident on the monochromator unit. Importantly, we have derived relations by which coherence properties and an average pulse duration may be determined in such experiments. Our analysis specifically took into consideration a finite resolution of the monochromator unit.

The results derived in this work are quite general and are not limited to the case of soft x-ray beamlines. They will be valid also for the XFEL hard x-ray beamlines as soon as intensity of the field delivered by the monochromator is expressed through its resolution function by a convolutional integral.

We should note here that the pulse duration determined in such experiments is strongly depending on the spectral properties of the x-ray beams. For example, if the spectrum is broadened due to frequency chirp effects this will lead to shorter pulse durations determined in HBT measurements. Such effects will need a special analysis.

The theoretical background provided here will be important for the statistical analysis of coherence properties of the XFEL beams at soft x-ray beamlines. This will provide the guideline for performing coherent experiments at the XFEL sources.

**Acknowledgements**

Authors acknowledge support of the project by E. Weckert and careful reading of the manuscript by R. Röhlsberger. The authors acknowledge discussions with Y.Y. Kim and with the European XFEL SCS Instrument beamline staff: A. Scherz, N. Gerasimova, and G. Mercurio who pointed out an importance of taking into account resolution function of the VLS monochromator in HBT analysis. I.A.V. acknowledge fruitful discussions with V. Kaganer on the subject of scattering from the gratings. This work was supported by the Helmholtz Associations Initiative Networking Fund (grant No. HRSF-0002) and the Russian Science Foundation (grant No. 18-41-06001).

**Appendix I**

Here we show how result in Eq. (95) is obtained. By substituting the cross-spectral density function in Eqs. (66, 67) in Eq. (93) one can transform integration in the numerator to (see also [27])

$$\iint e^{iq_1^D x_1 - iq_2^D x_2} W_{in}(x_1, x_2) dx_1 dx_2$$
$$= \iint exp\{-[a(x_1^2 + x_2^2) - 2bx_1 x_2]\} exp(iq_1^D x_1 - iq_2^D x_2) dx_1 dx_2 \quad (A1)$$
$$= \frac{\pi}{(a^2 - b^2)^{1/2}} exp\{-[\alpha[(q_1^D)^2 + (q_2^D)^2]] - 2\beta q_1^D \cdot q_2^D\},$$

where

$$\alpha = \frac{a}{4(a^2 - b^2)}, \beta = \frac{b}{4(a^2 - b^2)}. \quad (A2)$$

This result is obtained by the use of the known integral

$$\int e^{-at^2} e^{iqt} dt = \sqrt{\frac{\pi}{a}} e^{-q^2/4a}. \quad (A3)$$

In the denominator similar integration of Eq. (94) gives

$$S_D(q_i^D) = \int e^{iq_i^D x_1 - iq_i^D x_2} W_{in}(x_1, x_2) dx_1 dx_2 = \frac{\pi}{(a^2 - b^2)^{1/2}} e^{-2(\alpha-\beta)q_i^D}. \quad (A4)$$

Substituting results of integration in Eqs. (A1, A4) in Eq. (93) gives [34]

$$g_{in}(q_1^D, q_2^D) = e^{-\beta(q_2^D - q_1^D)^2}. \quad (A5)$$

**References**

1. J. Rossbach, J. R. Schneider, W. Wurth *10 years of pioneering X-ray science at the Free-Electron Laser FLASH at DESY,* Physics reports **808**, 1 - 74 (2019).

2. P. Emma, *et al., First lasing and operation of an ångstrom-wavelength free-electron laser,* Nat. Photon. **4**, 641–647 (2010).

3. Ishikawa T. *et al. A compact X-ray free-electron laser emitting in the sub-ångström region,* Nat. Photon. **6**, 540–544 (2012).

4. E. Allaria et al., *Two-stage seeded soft-X-ray free-electron laser,* Nat. Photon., **7** 913-918 (2013).

5. H.-S. Kang, *et al., Hard X-ray free-electron laser with femtosecond-scale timing jitter*, Nat. Photon. **11**, 708-713 (2017).

6. W. Decking et al., *A MHz-repetition-rate hard X-ray free-electron laser driven by a superconducting linear accelerator,* Nat. Photon. **14**, 391-397 (2020).

7. E. Prat *et al., A compact and cost-effective hard X-ray free-electron laser driven by a high-brightness and low-energy electron beam,* Nat. Photon. **14**, 748–754 (2020).

8. I. A. Vartanyants *et al., Coherence Properties of Individual Femtosecond Pulses of an X-Ray Free-Electron Laser*, Phys. Rev. Lett. **107**, 144801 (2011).

9. A. Singer *et al., Spatial and temporal coherence properties of single free-electron laser pulses,* Optics Express **20**(16) 17480 – 17495 (2012).

10. Ch. Gutt et al., *Single Shot Spatial and Temporal Coherence Properties of the SLAC Linac Coherent Light Source in the Hard X-Ray Regime,* Phys. Rev. Lett. **108**, 024801 (2012).

11. F. Lehmkühler *et al., Single Shot Coherence Properties of the Free-Electron Laser SACLA in the Hard X-ray Regime,* Sci. Rep. **4**, 5234 (2015).

12. Hanbury Brown, R. & Twiss, R. Q. *Correlation between Photons in two Coherent Beams of Light*, Nature **177**, 27–29 (1956)

13. Hanbury Brown, R. & Twiss, R. Q. A *Test of a New Type of Stellar Interferometer on Sirius*, Nature **178**, 1046–1048 (1956).

14. R. Loudon *The Quantum Theory of Light* (Oxford University Press, Oxford, 2000) 3rd edition.

15. R. J. Glauber, *The quantum theory of optical coherence,* Phys. Rev. **130**(6), 2529 (1963).

16. A. Singer, *et al., Hanbury Brown and Twiss interferometry at a free-electron laser*, Phys. Rev. Lett. **111**, 034802 (2013). Erratum, *ibid.* Phys. Rev. Lett. 117, 239903 (2016).

17. O. Yu. Gorobtsov, *et al., Statistical properties of a free-electron laser revealed by the Hanbury Brown and Twiss interferometry,* Phys. Rev. A **95**(2), 023843 (2017).

18. O. Yu. Gorobtsov *et al., Diffraction based Hanbury Brown and Twiss interferometry at a hard x-ray free-electron laser,* Sci. Rep. **8**, 2219 (2018).

19. O. Yu. Gorobtsov, *et al., Seeded X-ray free-electron laser generating radiation with laser statistical properties,* Nat. Comm. **9** 4498 (2018).

20. I. Inoue, T. Hara, Y. Inubushi, K. Tono, T. Inagaki, T. Katayama, Y. Amemiya, H. Tanaka, and M. Yabashi, *X-ray Hanbury Brown-Twiss interferometry for determination of ultrashort electron-bunch duration*, Phys. Rev. Acceler. and Beams **21**, 080704 (2018).

21. R. Khubbutdinov *et al., High spatial coherence and short pulse duration revealed by the Hanbury Brown and Twiss interferometry at the European XFEL,* (2021) (in preparation).

22. Y.Y. Kim et al., *Hanbury Brown and Twiss interferometry at PAL-XFEL*, (2021) (in preparation).

23. V.M. Kaganer, I. Petrov, L. Samoylova, *Resolution of a bent-crystal spectrometer for X-rays free-electron laser pulses: diamond vs. silicon,* Acta Cryst. A (2021) (in print) (see arXiv: 2009.08818 [physics.ins-det]).

24. S. Serkez, V. Kocharyan, E. Saldin, *Grating Monochromator for Soft X-ray Self-Seeding the European XFEL*, DESY Preprint 13-040 (2013).

25. M. Born, & E. Wolf, *Principles of Optics,* Cambridge University Press (Cambridge, 2019), 7$^{th}$ edition.

26. V. G. Kohn *The Method of Phase Retrieval of Complex Wavefield from Two Intensity Measurements Applicable to Hard X-rays*, Physics Scripta **56**, 14-19, (1997).

27. L. Mandel and E. Wolf, *Optical Coherence and Quantum Optics* (Cambridge University Press, 1995).

*28.* I. A. Vartanyants and A. Singer, *Coherence Properties of Third-Generation Synchrotron Sources and Free-Electron Lasers,* Chapter in book: *"Synchrotron Light Sources and Free-Electron-Lasers Accelerator Physics, Instrumentation and Science Applications"* Editors: E. J. Jaeschke, S. Khan, J. R. Schneider, J. B. Hastings (Springer Nature Switzerland AG Publishing; 2nd edition; 2020), pp. 987-1029.

29. D. Ferreira, R. Bachelard, W. Guerin, R. Kaiser, and M. Fouché, *Connecting field and intensity correlations: The Siegert relation and how to test it,* Am. J. Phys. **88**, 831-837 (2020).

30. J. W. Goodman, *Statistical Optics*, A Wiley-Interscience Publication (Wiley, 2000).

31. P. Pääkkönen, J. Turunen, P. Vahimaa, A. T. Friberg, and F. Wyrowski, *Partially coherent Gaussian pulses*, Opt. Commun. **204**, 53–58 (2002).

32. H. Lajunen, J. Tervo, and P. Vahimaa, *Theory of spatially and spectrally partially coherent pulses*, J. Opt. Soc. Am. A **22**, 1536 (2005).

33. The integration here was performed by the *Mathematica* software version 12.1.

34. We checked these results by performing integration with the *Mathematica* software version 12.1 and obtained the same result.

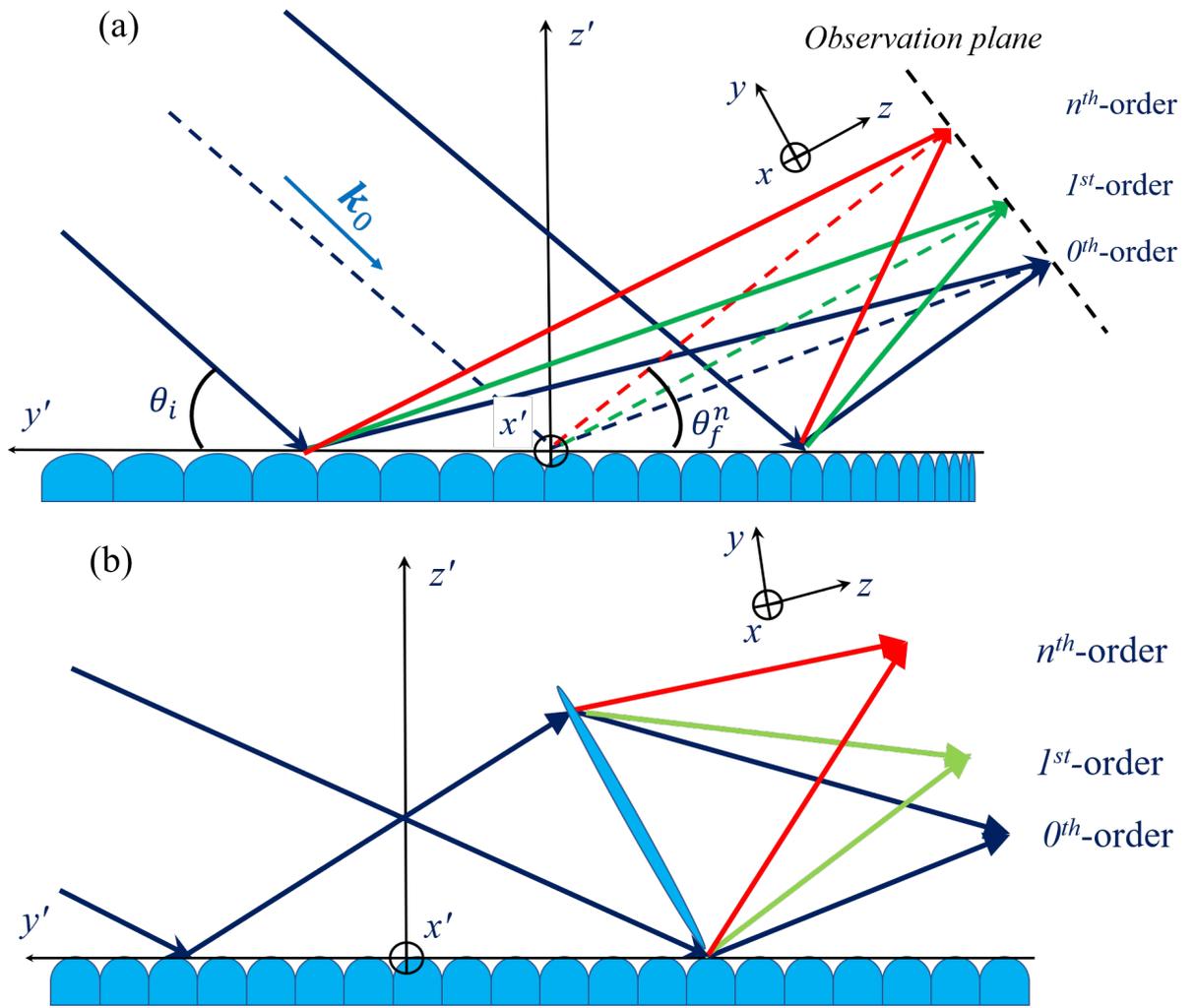

**Fig. 1.** (a) Scattering from the VLS grating is shown. Different spectral components of the incoming beam are focused at the focal distance $f$ forming the $0^{th}$, $1^{st}$, and $n^{th}$ orders of grating reflections. The coordinate systems for the grating and for the scattered beams is also shown. (b) Scattering from the VLS grating is substituted by the scattering on a grating with the constant period $d_0$ followed by a "virtual" lens with the focusing distance $f$.

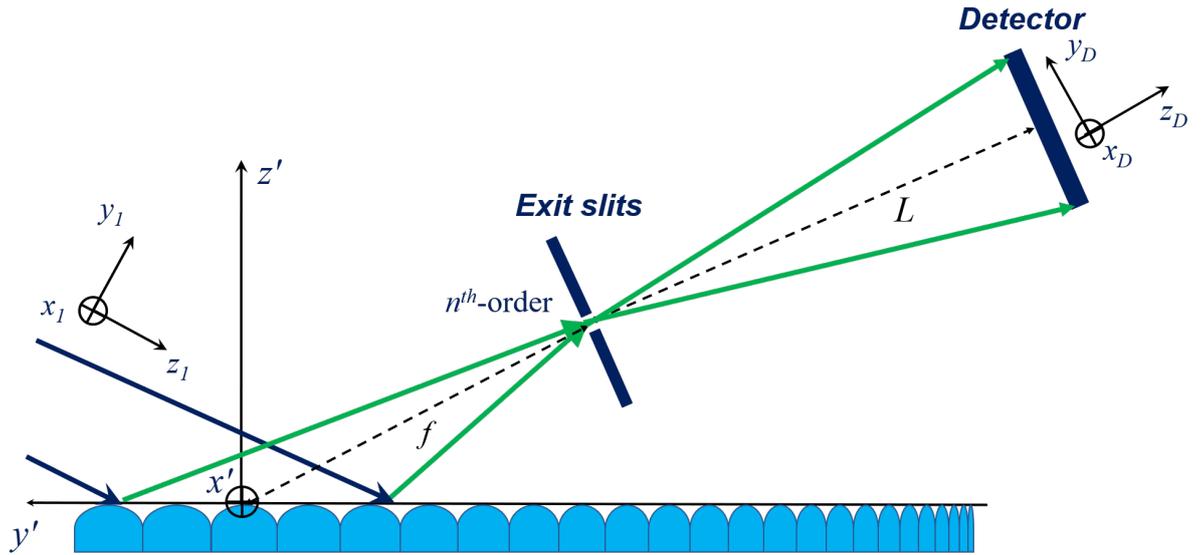

**Fig. 2.** Schematic layout of the HBT experiment performed at the XFEL facility. The incoming beam is scattered on the VLS grating. The exit slits are positioned at the focal plane of the VLS grating at the grating order $n$. The second-order spatial correlation measurements are performed by the pixelated detector positioned far from the monochromator unit to resolve spatial modes of the incoming X-ray pulses. The coordinate systems for the incoming beam, the grating, and detector are shown.